# The catalytic effect of water in basic hydrolysis of $CO_3^{2-}$ in hydrated clusters

Hang Xiao[1], Xiaoyang Shi[1], Yayun Zhang[1,4], Xiangbiao Liao[1], Feng Hao[1], Klaus S. Lackner[2] and Xi Chen[*1,3]

[1] *Columbia Nanomechanics Research Center, Department of Earth and Environmental Engineering, Columbia University, New York, NY 10027, USA*

[2] *School of Sustainable Engineering & Built Environment, Arizona State University, Tempe, AZ 85287-9309, USA*

[3] *SV Laboratory, School of Aerospace, Xi'an Jiaotong University, Xi'an 710049, China*

[4] *College of Power Engineering, Chongqing University, Chongqing 400030, China*

The hydration of ions in nanoscale hydrated clusters is ubiquitous and essential in many physical and chemical processes. Here we show that the hydrolysis reaction is strongly affected by relative humidity. The hydrolysis of $CO_3^{2-}$ with $n$ = 1-8 water molecules is investigated by *ab initio* method. For $n$ = 1-5 water molecules, all the reactants follow a stepwise pathway to the transition state. For $n$ = 6-8 water molecules, all the reactants undergo a direct proton transfer to the transition state with overall lower activation free energy. The activation free energy of the reaction is dramatically reduced from 10.4 to 2.4 kcal/mol as the number of water molecules increases from 1 to 6. Meanwhile, the degree of the hydrolysis of $CO_3^{2-}$ is significantly increased compared to the bulk water solution scenario. The incomplete hydration shells facilitate the hydrolysis of $CO_3^{2-}$ with few water molecules (especially for $n$ = 6) to be not only thermodynamically favorable but also kinetically favorable. This discovery provides valuable insights for designing efficient sorbents including that for $CO_2$ air capture.

## INTRODUCTION

The ability of an "inert" solvent to affect the kinetics and thermodynamics of a chemical reaction has been known for over 150 years [1]. Considerable efforts have been devoted to understand the role of solvents in bulk solutions [1–3]. Recently, the solvent effect in nanometer sized clusters or in nanoscale confinement has attracted increasing interest [4–8], due to its ubiquity and importance in varies biological and chemical processes [9–12]. Unlike the ion hydration in the bulk solution, the high ratio of ions to water molecules in nanoscale clusters and cavities could render the hydration shells incomplete. The hydrolysis of ions with these incomplete hydration shells could be significantly different from that in bulk water.

On the other hand, the development of efficient absorbents that can easily switch between absorption and desorption, has been of paramount importance for many processes. For example, direct air capture of $CO_2$ represents a promising carbon negative technology, and the major challenge of developing an efficient absorbent is not how to absorb $CO_2$, but how to release it with very low energy cost. This essentially requires a reversible chemical reaction that can be triggered by a simple environmental variable. Lackner et al. [13] discovered that an anionic exchange resin (IER) washed by carbonate solution can efficiently capture $CO_2$ from ambient air when it is dry, while release $CO_2$ when it is wet, as shown in Fig. 1. A better understanding of the hydrolysis of $CO_3^{2-}$ in hydrated clusters is of great importance for understanding of such a novel humidity-swing reaction with very low energy cost.

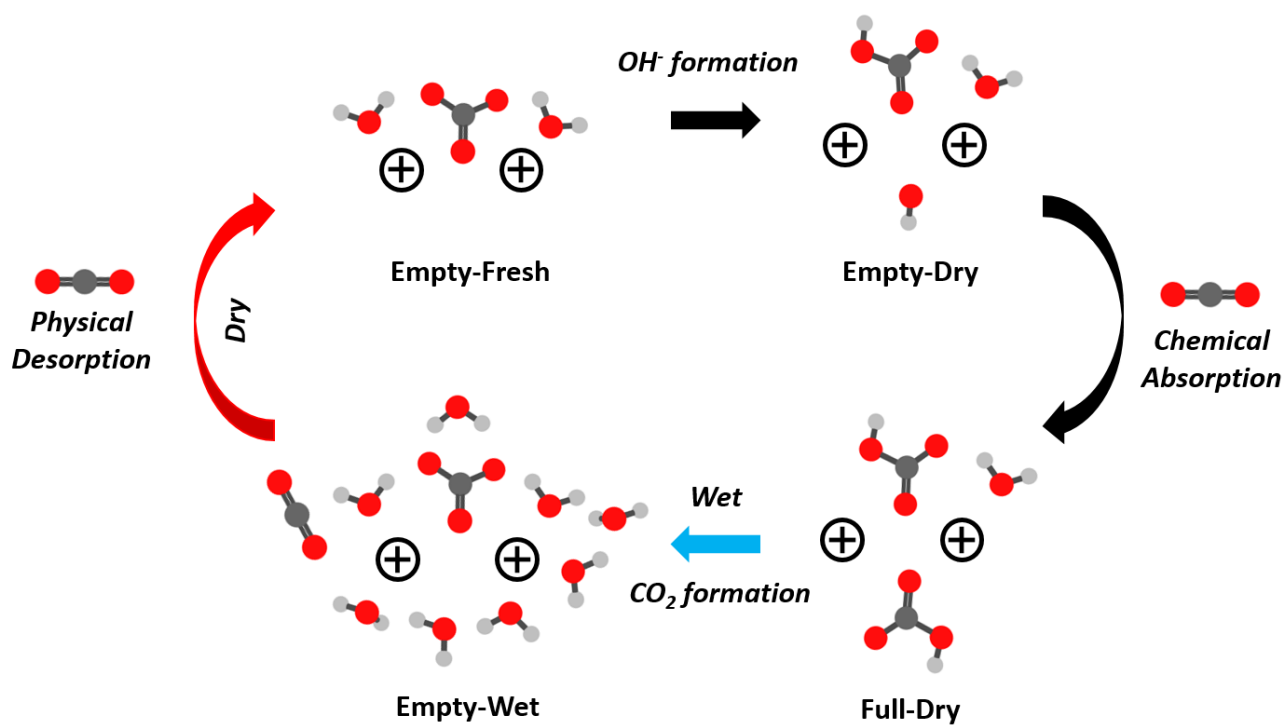

*Fig. 1: Humidity driven $CO_2$ absorption/desorption on IER. Empty-Fresh state: dry sorbent with only a few water molecules neighboring each carbonate ion. Empty-Dry state: $OH^-$ ion and $HCO_3^-$ ion are formed by hydrolysis of $CO_3^{2-}$ in the dry condition. Full-Dry state: the full-loaded sorbent in the dry condition. $OH^-$ formation and chemical absorption of $CO_2$ (Eq. 1-2) present the absorption process. Empty-Wet state: $CO_2$ regeneration in the wet condition (Eq. 3), which represents the physical desorption of $CO_2$.*

The absorption process (Eqs. 1-2, dry) and desorption process (Eq. 3, wet) are:

$$CO_3^{2-} + n\,H_2O \Leftrightarrow HCO_3^- + OH^- + (n-1)\,H_2O \tag{1}$$

$$OH^- + CO_2 \Leftrightarrow HCO_3^- \tag{2}$$

$$2HCO_3^-(s) \Leftrightarrow CO_3^{2-}(aq) + H_2O + CO_2(g) \tag{3}$$

Our recent atomistic study [14,15] showed that the free energy of $CO_3^{2-}$ hydrolysis (Eq. 1) it is negative when the number of participating water molecules, $n$, is smaller than about 10. That is, the chemical reaction shifts to the right hand side (which is against the mass action law) when only a few water molecules surround each carbonate ion, rendering the material ready for $CO_2$ absorption through Eq. 2. With a large number of water molecules presenting, Eq. 1 swings to the left hand side like that in a bulk environment. When the fully loaded absorbent (after Eqs. 1-2) is subsequently placed in a wet environment, Eq. 3 releases $CO_2$ in gas phase, completing the absorption-desorption cycle and direct air capture of carbon dioxide.

While the thermodynamic characteristics of this sorbent have been investigated [14,15], the kinetic counterpart still remains to be clarified for such a humidity-swing process. The kinetic information of a chemical reaction (e.g. activation free energy) is particularly important, since only chemical reactions with low activation free energy are able to proceed at a reasonable rate. Therefore, the activation free energy of the hydrolysis of $CO_3^{2-}$ in hydrated clusters of different sizes needs to be investigated using quantum chemical calculations.

In this paper, the reaction pathways of the hydrolysis of $CO_3^{2-}$ with $n = 1$-8 water molecules (Eq. 1) are investigated theoretically. We elucidate how water molecules modulate the reaction pathways of $CO_3^{2-}$ hydrolysis and its underlying mechanism. It is found that the activation free energy of the $CO_3^{2-}$ hydrolysis reaction varies with the number of water molecules. In addition, nano-confinement is perhaps not a necessity for the humidity driven $CO_2$ air capture. These kinetic insights for $CO_3^{2-}$ hydrolysis in hydrated clusters will pave the way for designing efficient $CO_2$ air capture sorbents.

## RESULTS

**Hydrolysis reaction with *n* = 1-5**

Herein, we compare the hydrolysis of $CO_3^{2-}$ with different number of water molecules ($n$ = 1-5 in Eq. 1). The optimized structures and the corresponding relative free energy profiles of reaction pathways are presented in Fig. 2(a) and in Fig. 2(b), respectively. In order to balance the charge of carbonate anion, two mobile sodium cations are introduced into the system. Note that only the most promising reaction pathways (with lowest activation free energy) are presented, due to the increasing number of possible reaction pathways as *n* increases.

For the reaction with only 1 water molecule, the reaction follows a two-stage route to the product: (1) the $H_2O$ molecule migrates to a position where the proton transfer to the neighbor oxygen atom is energetically favorable, forming the intermediates denoted as I-1a. (2) Followed by the proton transfer through the transition state TS-1 to the product P-1.

For the reaction with 2-5 $H_2O$ molecules, a three-step route to the product is likely. The first step is the same with the reaction with 1 $H_2O$ molecule. However, through the transition states (TS-2, TS-3, TS-4 and TS-5), the proton transfer reactions leads to intermediates (I-2b, I-3b, I-4b and I-5b), followed by the migration of ions and $H_2O$ molecules to form the final product. One notes that the migrations of $H_2O$ molecules and ions proceed with little or no barrier, due to the absence of chemical reaction.

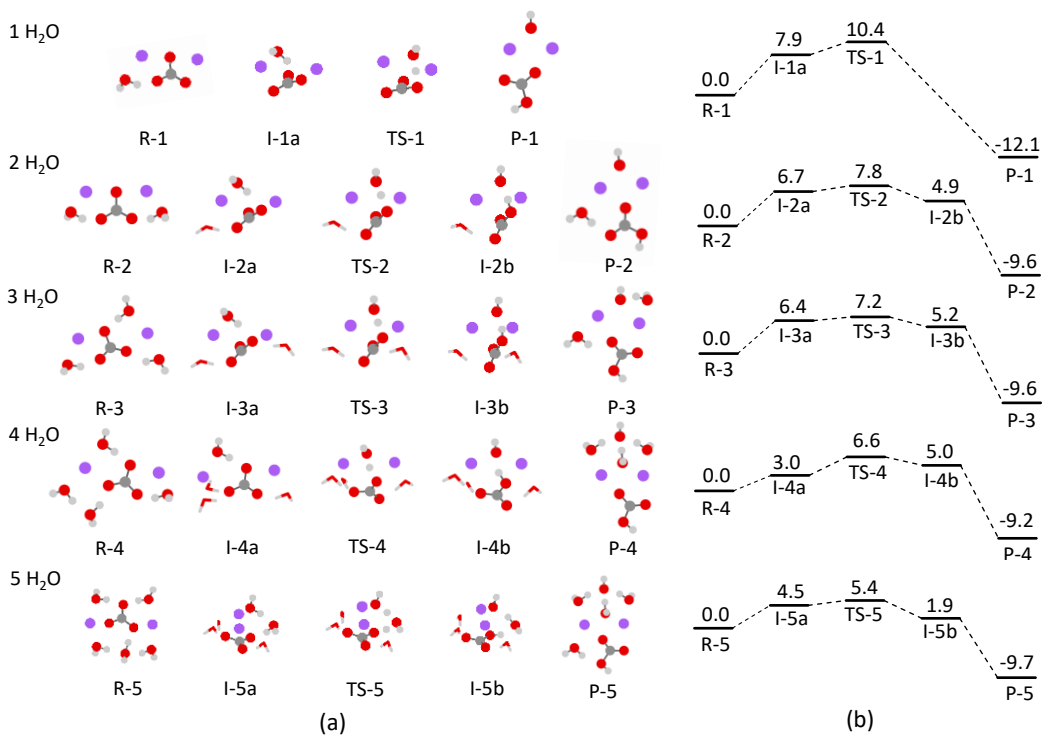

*Fig. 2. (a) Reactants, intermediates, products and transition states for the reaction in Eq. 1 (n =1-5). (b) Relative free energy profiles (in kcal/mol) for the hydration of $CO_3^{2-}$ with 1-5 water molecules. For transition states and intermediate states, the sodium ions, carbonate ions, bicarbonate ions, hydroxyl ions and the water molecules directly involved in reaction are visualized with the ball-and-stick model, while the water molecules do not directly take part in the reaction are visualized with the tube model. For reactants and products, all species are visualized with the ball-and-stick model. The same visualization protocol is adopted in Fig. 4.*

The activation free energy decreases as *n* grows from 1 to 5, as shown in Fig. 3. The reaction free energy increases by 2.5 kcal/mol as *n* grows from 1 to 2. However, the reaction free energy remains more or less the same as *n* rises from 2 to 5.

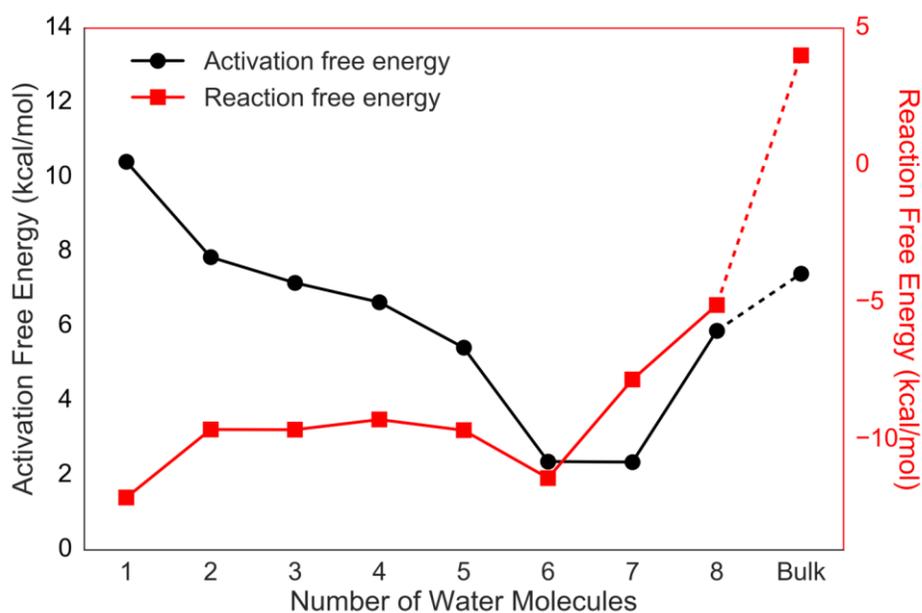

*Fig. 3. The activation free energy (in black) of Eq. 1 as a function of the number of $H_2O$ molecules; the reaction free energy (in red) of Eq. 1 as a function of the number of $H_2O$ molecules. The activation free energy and reaction free energy in bulk water are calculated with 8 explicit $H_2O$ molecules using the SMD continuum solvation model [16].*

**Hydrolysis reaction with *n* = 6-8**

Here, we consider the hydrolysis of $CO_3^{2-}$ with $n$ = 6-8 water molecules for comparison. The optimized structures of species involved in the hydrolysis reactions and the corresponding relative free energy profiles of reaction pathways are shown in Fig. 4(a) and in Fig. 4(b), respectively.

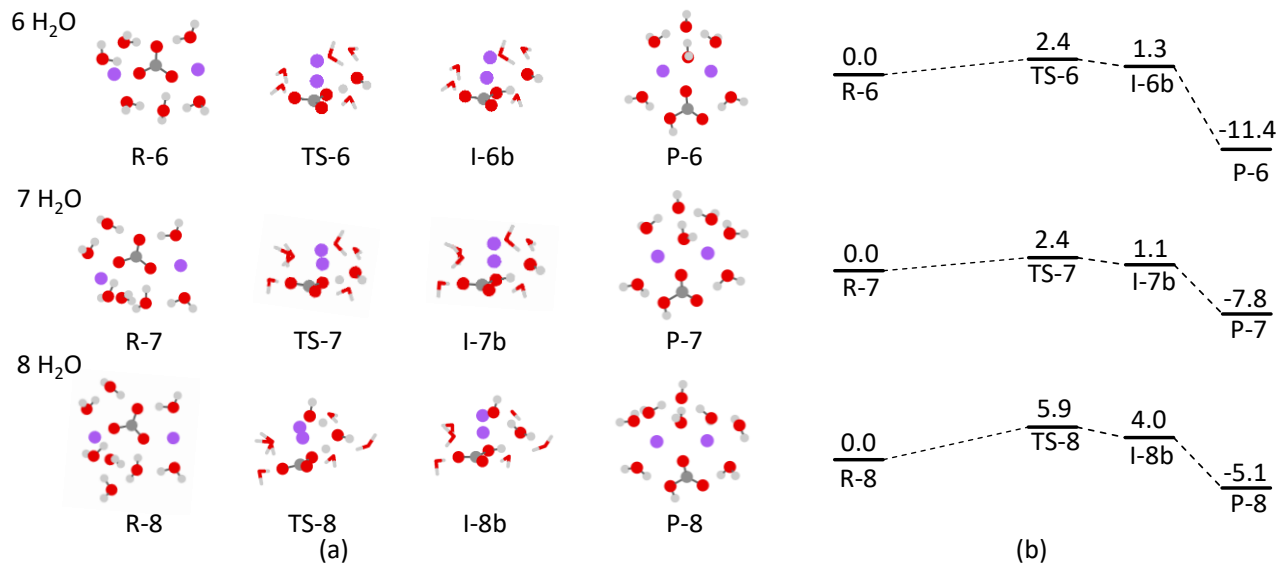

Fig. 4. (a) Reactants, intermediates, products and transition states for the reaction in Eq. 1 (n = 6-8). (b) Relative free energy profiles (in kcal/mol) for the hydration of $CO_3^{2-}$ with 6-8 water molecules.

For $n$ = 1-5 water molecules, all the reactants follow a stepwise pathway to the transition state through the intermediates (I-1a, I-2a, I-3a, I-4a and I-5a). However, for the reactions with 6-8 $H_2O$ molecules, the reactants (R-6, R-7 and R-8) undergo a proton transfer directly leading to the transition state, with overall lower activation free energy, as shown in Fig. 3.

For the reactions with $n$ = 6 and $n$ = 7, the single proton transfer occurs, i.e. only one water molecule is involved in the proton transfer reaction. While for $n$ = 5 and $n$ = 8, the water mediated double proton transfer is observed. Counterintuitively, the single proton transfers with $n$ = 6 and $n$ = 7 have a much lower activation free energy than the water-mediated proton transfers with $n$ = 5 and $n$ = 8, since water-mediated proton transfer is known to lower the energy barrier of proton transfer reactions [17–19].

**Comparison with the hydrolysis reaction in the bulk water ($n \gg 1$)**

The activation free energy and reaction free energy in bulk water ($n \gg 1$) are calculated with 8 explicit $H_2O$ molecules in a water dielectric using the SMD continuum solvation model [7], as shown in Fig. 3. The reaction free energy in bulk water calculated is 4.0 kcal/mol, a good agreement with the experimental value (5.0 kcal/mol) at the ambient condition [8]. The activation free energy in bulk water is slightly higher than the barrier in reaction with 8 water molecules.

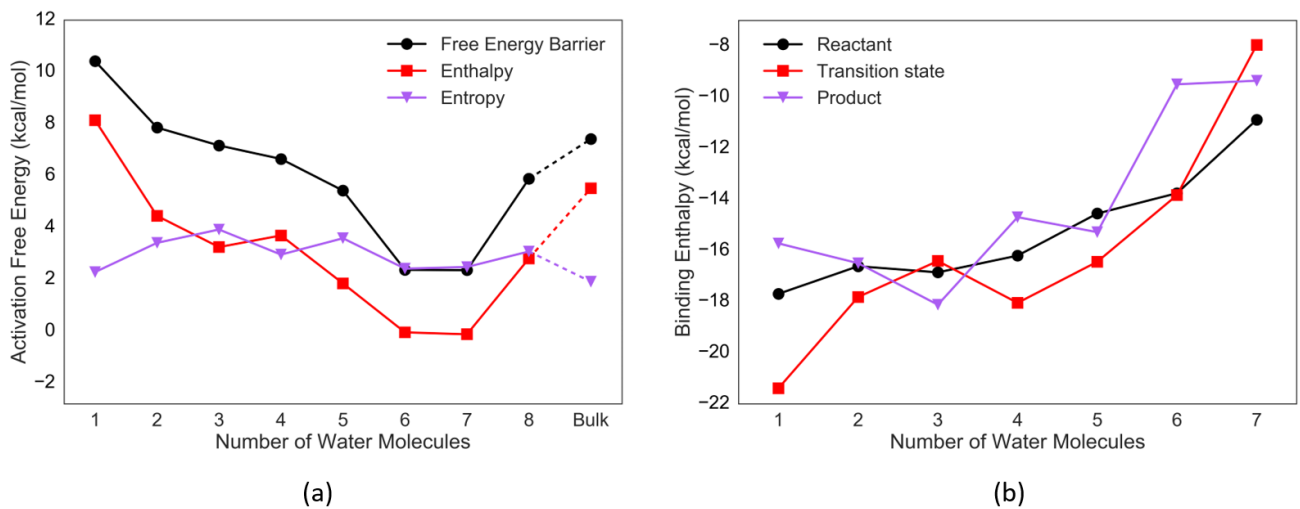

(a)  (b)

*Fig. 5. (a) The enthalpic (in red) and entropic (in purple) components of the activation free energy of Eq. 1 as a function of the number of $H_2O$ molecules. (b) Binding enthalpy of adding one $H_2O$ to reactants (in black), transition states (in red) and products (in purple) of the reactions with n water molecules. It can be calculated by $\Delta H_n = H_{X_{n+1}} - H_{X_n} - H_{H_2O}$.*

**The driving force of the change in activation free energy.**

To understand the driving force of the change in activation free energy with different number of water molecules, we decompose the activation free energy into enthalpic and entropic components, as shown in Fig. 5 (a). Clearly, the change of activation free energy is dominated by the change in its enthalpic component, which is discussed in detail in the following. The binding enthalpy of adding one $H_2O$ to reactants, transition states and products of the reactions with *n* water molecules can be calculated by $\Delta H_n = H_{X_{n+1}} - H_{X_n} - H_{H_2O}$, shown in Fig. 5 (b). For *n* = 1-5, the binding enthalpy of an extra $H_2O$ to transition states are generally lower than that of reactants. That is, water molecules adding to the system tend to stabilize the transition state structure more than the reactants, resulting in the drop of activation free barrier for *n* =1-6. For *n* = 7, the binding enthalpy of an extra $H_2O$ to transition state is much higher than that of reactant, which means that the extra $H_2O$ molecule tends to stabilize the reactant more than it does to the transition state. As a result, the activation barrier increases abruptly as *n* grows from 7 to 8.

**Implication to the humidity driven $CO_2$ air capture.**

The binding enthalpy of adding one $H_2O$ to reactants generally increases as the cluster size increases, as shown in Fig. 5 (b). That is, water molecules bind more firmly with smaller ion clusters. As a result, for two different scenarios of the adsorption of $H_2O$ on $CO_3^{2-}$ anchored on the surface of a porous material at low humidity (Fig. 6), the uniformly adsorption case is enthalpically favorable. Obviously, the uniformly adsorption case is entropically favored as well. Hence the water molecules tend to be more or less uniformly clustered around $CO_3^{2-}$ ions anchored on the surface of a material at low humidity – although such a reaction system is nanometer-sized, it does not require nano-confinement. In these nanometer-sized hydrated clusters, the $CO_3^{2-}$ hydrolysis reactions are able to spontaneously generate $OH^-$ ions that are ready to capture $CO_2$ from air at room temperature at low humidity. The employment of a nanoporous material helps to maximize the surface area (and hence anchored $CO_3^{2-}$ density) for higher efficiency air capture of $CO_2$ (as long as the carbonate ions are anchored uniformly and firmly), but the confinement from nanoporous is not a required condition for the humidity-swing hydrolysis reaction, contrary to the former proposal [14].

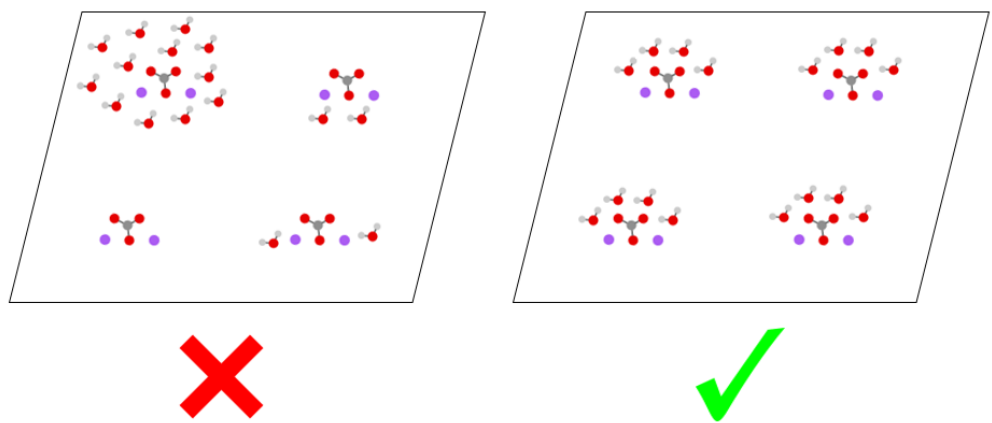

Fig. 6. Two different scenarios of the adsorption of $H_2O$ on $CO_3^{2-}$ anchored on the surface of a porous material at low humidity.

## DISCUSSIONS

The reaction free energy determines the equilibrium point of a chemical reaction, while the activation free energy determines the reaction kinetics. As $n$ increases, the activation free energy of $CO_3^{2-}$ hydrolysis firstly monotonically decreases from 10.4 kcal/mol ($n = 1$) to the minimum value 2.4 kcal/mol ($n = 6$ and $n = 7$), then increases again to 7.4 kcal/mol ($n \gg 1$), as shown in Fig. 3. The incomplete hydration shells involved in reactions with $n = 6$ and $n = 7$ render the $CO_3^{2-}$ hydrolysis kinetically favorable. Note that the reaction free energies in reactions with $n = 1-8$ water molecules are actually negative, indicating the incomplete hydration shells also render the $CO_3^{2-}$ hydrolysis (Eq. 1) thermodynamically favorable.

Given the reaction free energy for reaction with $n = 7$ is much higher than the reaction with $n = 6$, the incomplete hydration shell with 6 water molecules appears to be the best choice for the system considered in this study, in terms of promoting the $CO_3^{2-}$ hydrolysis reaction both kinetically and thermodynamically. The $CO_3^{2-}$ hydrolysis is the rate limiting step in $CO_2$ air capture processes developed by Lackner et al [13]. Our result is able to provide valuable insights to designing efficient $CO_2$ air-capture sorbents for applications in environment with different humidity (e.g. designing $CO_3^{2-}$ anchored nanoporous materials that facilitate the formation of incomplete hydration shell of $CO_3^{2-}$ within a specific range of humidity that corresponds to 1:6 ratio of carbonate ion vs. water molecules). This catalytic effect of water molecules is not limited to the hydrolysis of $CO_3^{2-}$ with incomplete hydration shells. It is expected that incomplete hydration shells will have similar effects on the hydrolysis of different types of salts: as remarked in recently in the thermodynamics study our work [15] the hydrolysis reactions of other basic salts are also affected by humidity, and their kinetics may also be studied using the framework proposed in this paper to optimize the design of efficient absorbents.

## METHODS

Global-minimum structural searches for stable reactants and products for the reaction in Eq. 1 were carried out using the Minima Hopping algorithm [20] implemented in CP2K [21] at the PBE-D3/DZVP level [22–24]. Subsequently, the stable reactants and products were fully optimized at the B3LYP /6-311+G(2d,2p) level [6] with D3 version of Grimme's dispersion correction with Becke-Johnson damping [7], with the Gaussian 09 package [24]. Geometries of all transition states and intermediates were fully optimized at the same level. To account for the effects of an aqueous environment, the activation free energy and reaction free energy in bulk water are calculated with 8 explicit $H_2O$ molecules in a water dielectric using the SMD continuum solvation model [16]. The reaction free energy in bulk water calculated is 4.0 kcal/mol, which agrees well with the experimental value (5.0 kcal/mol) at the ambient condition [25]. Frequency calculations have been carried out to check for the nature of the various stationary points and transition states, which were also used for the computation of zero-point, thermal and entropy contributions to free energy at 298 K. The correlation between the stable structures and the transition states is further verified by the intrinsic reaction coordinate calculations.

# ACKNOWLEDGEMENTS

X.C. acknowledges the support from the National Natural Science Foundation of China (11172231 and 11372241), ARPA-E (DE-AR0000396) and AFOSR (FA9550-12-1-0159)

.